\documentclass[10pt,conference]{IEEEtran}
\IEEEoverridecommandlockouts

\usepackage{cite}
\usepackage{amsmath,amssymb,amsfonts}
\usepackage[linesnumbered,ruled,vlined]{algorithm2e}
\usepackage{graphicx}
\usepackage{textcomp}
\usepackage{xcolor}
\usepackage{subfiles}
\usepackage{multirow}
\usepackage{makecell}
\usepackage{caption}
\usepackage{hhline}

\def\BibTeX{{\rm B\kern-.05em{\sc i\kern-.025em b}\kern-.08em
    T\kern-.1667em\lower.7ex\hbox{E}\kern-.125emX}}

\begin{document}

\title{
CaMDN: Enhancing Cache Efficiency for Multi-tenant DNNs on Integrated NPUs
\thanks{This work was supported by the National Key R\&D Program of China under Grant No. 2023YFB4503100, the National Natural Science Foundation of China under Grant No. 62272026 and No.62104014, the Fundamental Research Funds for the Central Universities, State Key Laboratory of Complex \& Critical Software Environment under Grant No. CCSE-2024ZX-10, and Beijing Advanced Innovation Center for Future Blockchain and Privacy Computing.}
\thanks{\textsuperscript{*} Liang Wang and Limin Xiao are corresponding authors of this paper.}
}

\author{
\IEEEauthorblockN{
Tianhao Cai\textsuperscript{1,2}, 
Liang Wang\textsuperscript{1,2,*}, 
Limin Xiao\textsuperscript{1,2,*}, 
Meng Han\textsuperscript{3}, 
Zeyu Wang\textsuperscript{1,2}, 
Lin Sun\textsuperscript{4} and
Xiaojian Liao\textsuperscript{1,2}
}
\IEEEauthorblockA{\textsuperscript{1}
State Key Laboratory of CCSE, Beihang University, Beijing, China
} 
\IEEEauthorblockA{\textsuperscript{2}
School of Computer Science and Engineering, Beihang University, Beijing, China
} 
\IEEEauthorblockA{\textsuperscript{3}
School of Integrated Circuits, Tsinghua University, Beijing, China
}
\IEEEauthorblockA{\textsuperscript{4}
Jiangsu Shuguang Optoelectric Co., Ltd., Yangzhou, China
}

\IEEEauthorblockA{
caitianhao@buaa.edu.cn, lwang20@buaa.edu.cn, xiaolm@buaa.edu.cn
}
}

\maketitle

\begin{abstract}
With the rapid development of DNN applications, multi-tenant execution, where multiple DNNs are co-located on a single SoC, is becoming a prevailing trend. Although many methods are proposed in prior works to improve multi-tenant performance, the impact of shared cache is not well studied. This paper proposes CaMDN, an architecture-scheduling co-design to enhance cache efficiency for multi-tenant DNNs on integrated NPUs. Specifically, a lightweight architecture is proposed to support model-exclusive, NPU-controlled regions inside shared cache to eliminate unexpected cache contention. Moreover, a cache scheduling method is proposed to improve shared cache utilization. In particular, it includes a cache-aware mapping method for adaptability to the varying available cache capacity and a dynamic allocation algorithm to adjust the usage among co-located DNNs at runtime. Compared to prior works, CaMDN reduces the memory access by 33.4\% on average and achieves a model speedup of up to 2.56$\times$ (1.88$\times$ on average).
\end{abstract}


\section{Introduction}

Deep Neural Networks (DNNs) are increasingly pervasive in various applications. This brings an inevitable demand for efficient multi-tenant DNN execution, where many models are co-located on a single SoC to improve the performance of comprehensive DNN applications. Therefore, modern SoCs tend to integrate Neural Processing Units (NPUs) alongside CPUs to support trending multi-DNN applications \cite{hotchip-qualcomm, hotchip-amdzen5, hotchip-intel}. 

The key challenge of co-locating DNNs is the intense contention among running tasks over all kinds of shared resources. Methods are proposed to schedule different resources among models through either temporal interleaving\cite{layerweaver, prema, aimt} or spatial partitioning\cite{magma, moca, planaria, dream, relmas, hda, aurora, veltair}. However, these methods mainly focus on memory bandwidth\cite{magma, moca}, NPU cores\cite{planaria, dream}, or both\cite{relmas, hda, aurora}. The performance impact of shared cache with multi-tenant DNNs is still not well studied. 

Cache efficiency can be severely compromised in multi-tenancy due to the contention among running programs. Our experiment shows that cache hit rate drops by up to 59.7\%, and memory access increases by up to 64.1\% with 32 co-located DNNs, as shown in Section \ref{sec_moti_expr}. Since multi-tenant DNNs are generally memory-bound, the increase in memory access directly results in severe model slowdown. According to our statistical analyses, a large portion of data in multi-tenant DNNs is either non-reusable or reused over long distances on shared cache, which is a primary cause of the inefficiency.

Prior works manage to improve cache efficiency with different types of multi-tenant workloads other than DNNs, such as CPU workloads\cite{qureshi2006utility, sanchez2011vantage, herdrich2016cache, park2019copart} and accelerator workloads\cite{bufferincache, bufferinnuca, cohmeleon}. However, the performance improvements brought by these methods are limited due to either the need to keep cache transparent to workloads, or lack of consideration on special characteristics of trending multi-tenant DNNs. 

For multi-tenant DNNs, both architecture and scheduling require specialized optimization to improve shared cache efficiency. In terms of architecture, cache is required to be able to bypass non-reusable data and retain data with long reuse distances. In terms of scheduling, model mapping is required to be aware that cache will be shared among multiple models, and resource allocation is required to dynamically respond to changes of cache usage during multi-tenant execution.

Therefore, this paper proposes CaMDN, an architecture-scheduling co-design to enhance cache efficiency for multi-tenant DNNs on integrated NPUs. Specifically, CaMDN introduces a lightweight architecture to support model-exclusive, NPU-controlled regions inside shared cache. This allows NPUs to manage data storage and replacement on cache and eliminates unexpected contention among models. To fully utilize the architectural advantages, CaMDN provides a cache scheduling method including cache-aware mapping and dynamic cache allocation. In particular, cache-aware mapping generates multiple mapping candidates that target different cache usage levels to provide adaptability to the varying available capacity. Dynamic cache allocation adjusts cache usage among co-located DNNs at runtime. It responds quickly to changes of cache usage to improve cache utilization. In summary, this paper makes the following main contributions:
\begin{itemize}
    \item We propose a lightweight architecture to support model-exclusive, NPU-controlled regions inside shared cache.
    \item We propose a cache scheduling method that contains cache-aware DNN mapping and dynamic cache allocation to improve shared cache utilization.
    \item We demonstrate that, compared to prior works, CaMDN reduces the memory access by 33.4\% and achieves a model speedup of up to 2.56$\times$ (1.88$\times$ on average).
\end{itemize}

\section{Background and Motivation}

\subsection{Multi-tenant DNN Execution}

As DNNs increasingly pervade various applications, multi-tenant execution becomes an inevitable requirement for multiple reasons. First, complex applications such as AR/VR are likely to apply more than one DNN. Second, multiple DNN applications may run on a single device like personal computers and edge devices\cite{omniboost}. Third, cloud servers need to support multi-tenancy for modern AI services\cite{tpuv4i}. 

\begin{figure}[t]
\centering
\includegraphics[scale=0.9]{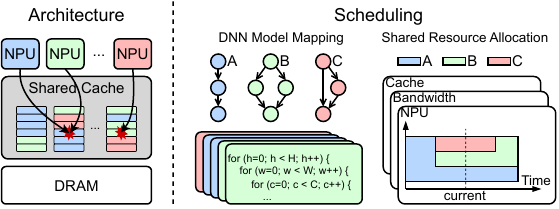}
\vspace{-5pt}
\caption{An overview of architecture and scheduling for multi-tenant DNN execution on integrated NPUs. Cache contention occurs between co-located DNNs.}
\vspace{-15pt}
\label{fig_overview}
\end{figure}

Figure \ref{fig_overview} illustrates an overview of architecture and scheduling for multi-tenant DNN execution on a single SoC. Architecturally, to meet the trending demand for multi-tenant DNNs, the chip industry has started to integrate NPUs alongside CPUs on their SoCs\cite{hotchip-qualcomm, hotchip-amdzen5, hotchip-intel}. Such architectures leverage the benefits of shared memory subsystem, including higher interconnect bandwidth, reduced host-device data movement, and additional on-chip storage provided by shared cache. 

Scheduling refers to the process of deploying DNN models on target hardware. In particular, model mapping is to generate hardware-specific programs from high-level model descriptions. Mapping is demonstrated to have a significant impact on model performance, and many optimizations are proposed in prior works\cite{timeloop, cosa, dmazerunner}. At runtime, shared resources, such as NPUs and bandwidth, are allocated to models according to their requirements as well as target deadlines.

\subsection{Scheduling Multi-tenant DNNs}

Recently, methods are proposed in prior works to schedule multi-tenant DNNs on integrated NPUs. Some of them use temporal interleaving to execute tasks in turn\cite{layerweaver, aimt, prema}. These methods suffer from under-utilization as resources scale richer. Others use spatial partitioning to fully utilize shared resources. We categorize these methods according to the resources they mainly focus on. 
\subsubsection{Bandwidth} MAGMA\cite{magma} and MoCA\cite{moca} allocate memory bandwidth usage among co-located DNNs according to their memory access requirements.
\subsubsection{Compute} Planaria\cite{planaria} manages to co-locate multiple models on a single large systolic array-based NPU. DREAM\cite{dream} dynamically schedules multiple NPUs for multi-tenant DNNs in real-time scenarios.
\subsubsection{Bandwidth \& Compute} RELMAS\cite{relmas} considers bandwidth contention for NPU scheduling but does not employ any hardware constraints on bandwidth usage. HDA\cite{hda} and AuRORA\cite{aurora} leverage static and dynamic scheduling methods, respectively, to co-allocate bandwidth and NPUs. 

Prior works above manage to schedule memory bandwidth and/or NPUs to achieve higher performance. However, they fail to consider the performance impact of shared cache with multi-tenant DNNs. Veltair\cite{veltair} uses cache performance counters for contention monitoring, but does not utilize any hardware method to address the issue. Additionally, Veltair targets server-level many-core CPUs rather than NPUs.

\begin{figure}[t]
\centering
\includegraphics[scale=0.5]{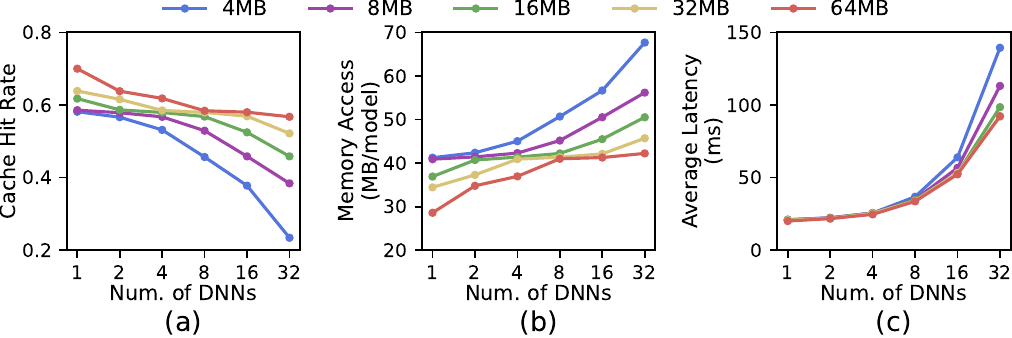}
\vspace{-5pt}
\caption{Cache hit rate, memory access and average latency with different settings on number of DNNs and cache capacity. }
\vspace{-5pt}
\label{fig_motivation_expr}
\end{figure}

\begin{figure}[t]
\centering
\includegraphics[scale=0.49]{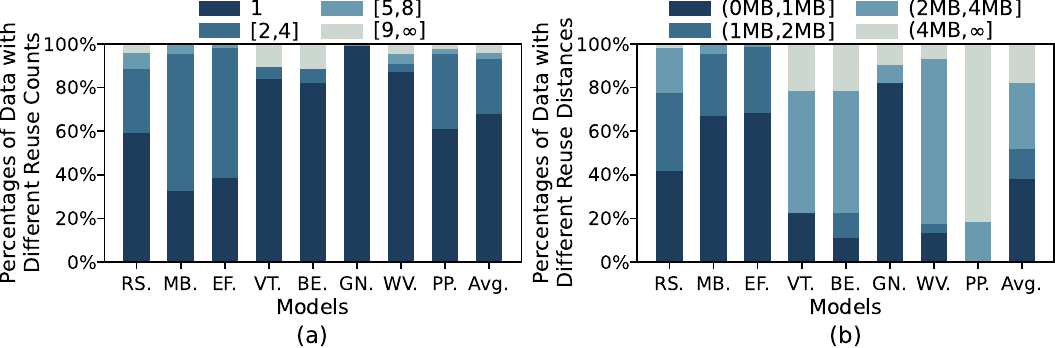}
\vspace{-5pt}
\caption{Percentages of data with different reuse counts and reuse distances on shared cache in benchmark DNN models.}
\vspace{-15pt}
\label{fig_motivation_expr_reuse}
\end{figure}

\subsection{Cache Inefficiency with Multi-tenant DNNs}
\label{sec_moti_expr}
To understand the performance impact of shared cache with multi-tenant DNNs, we conduct an experiment that randomly dispatches different DNN models on an NPU-integrated SoC. The experiment is performed with different settings on the number of co-located models and shared cache capacity. The detailed experimental setup is described in Section \ref{ref_expr_setup}. Figure \ref{fig_motivation_expr} shows the results of the experiment. The cache hit rate drops by 18.9\% to 59.7\%, and the memory access increases by 32.7\% to 64.1\% as the number of DNNs reaches 32. This is due to the severe cache contention among models, which occurs when a process frequently replaces cache lines belonging to other co-running processes, as illustrated in the left part of Figure \ref{fig_overview}. As a result, the average model latency increases by 3.46$\times$ to 5.65$\times$. Since multi-tenant DNNs are generally memory-bound, the increase in memory access directly contributes to the slowdown.

To further understand how DNN workload characteristics incur the inefficiency of shared cache, we perform the following statistical analyses on benchmark DNNs in Table \ref{tab_benchmark}. Figure \ref{fig_motivation_expr_reuse}(a) shows the percentages of data with different reuse counts. A reuse count refers to the expected number of repeated cache accesses to a piece of data. On average, 68.0\% of data have no future reuse but still occupy cache space, which results in a large amount of reusable data being early replaced off-chip. Figure \ref{fig_motivation_expr_reuse}(b) shows the percentages of intermediate data with different reuse distances when accessed across DNN layers. A reuse distance refers to the amount of accesses to other data that occur between two repeated accesses to the same piece of data. On average, 61.8\% of intermediate data have reuse distances more than 1MB, and 47.9\% of them more than 2MB. Such large reuse distances make it difficult for transparent cache to retain these data for future reuse, especially when multiple models are competing for cache space. In summary, a large portion of data in DNNs is either non-reusable or reused over long distances. Shared cache shows evident inefficiency to handle such data, therefore significantly degrading the performance of multi-tenant DNN execution.

\subsection{Enhancing Cache Efficiency}
Prior works manage to enhance cache efficiency for different multi-tenant workloads other than DNNs. For CPU workloads, cache partitioning and related techniques are proposed to limit the cache usage of each process\cite{qureshi2006utility, sanchez2011vantage, herdrich2016cache, park2019copart}. However, these methods offer limited performance improvement, since cache must be hardware-managed and transparent to CPU workloads. For accelerator workloads, methods are purposed that either instantiate accelerator-controlled buffers inside shared cache \cite{bufferincache, bufferinnuca}, or dynamically change cache coherence mode for each workload at runtime \cite{cohmeleon}. However, these methods lack consideration for the special characteristics of multi-tenant DNNs.

For multi-tenant DNNs, both architecture and scheduling require specialized optimization to improve shared cache efficiency. In terms of architecture, cache is required to be able to bypass non-reusable data and retain data with long reuse distances. In terms of scheduling, mapping is required to be aware that cache will be shared among multiple models, and resource allocation is required to dynamically respond to changes of cache usage during multi-tenant execution. Therefore, this paper proposes CaMDN, an architecture-scheduling co-design to enhance cache efficiency for multi-tenant DNNs on integrated NPUs. Specifically, CaMDN contains a lightweight architecture to support model-exclusive, NPU-controlled regions inside shared cache, a cache-aware DNN mapping method for adaptability to the varying available cache capacity, and a dynamic cache allocation algorithm to adjust the cache usage among co-located DNNs at runtime.

\begin{figure}[t]
\centering
\includegraphics[scale=0.79]{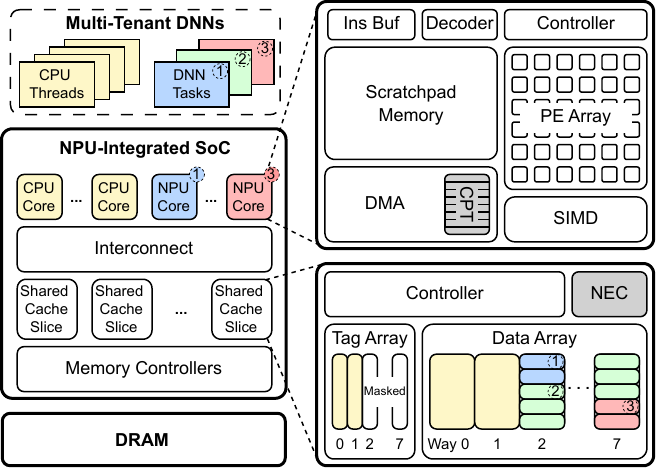}
\caption{The overview of CaMDN architecture. CaMDN installs an NEC in each shared cache slice and a CPT in each NPU.}
\label{fig_overall_arch}
\end{figure}

\section{CaMDN System}

\subsection{Overview}
The architecture-scheduling co-design of CaMDN contains the following components. (1) An NPU-controlled cache architecture that supports model-exclusive, NPU-controlled regions inside shared cache by adding an NPU-exclusive controller (NEC) in each cache slice and a cache page table (CPT) in each NPU (Section \ref{sec_hardware}). (2) A cache-aware mapping method that provides adaptability to varying available cache size by generating multiple mapping candidates that target different cache usage levels (Section \ref{sec_mapping}). (3) A dynamic cache allocation algorithm that adjusts cache usage among models by predicting near-future cache usage and selecting mapping candidates accordingly (Section \ref{sec_scheduling}).

\begin{figure}[t]
\centering
\includegraphics[scale=0.85]{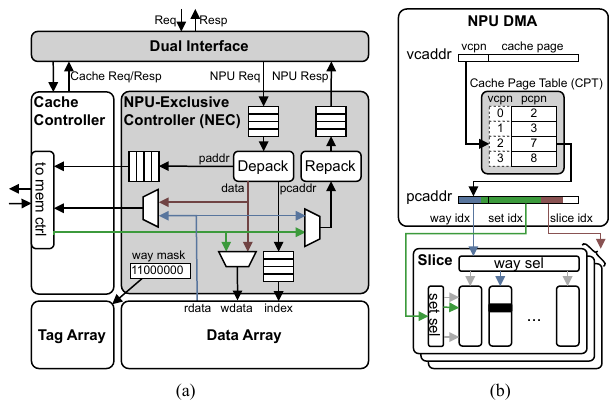}
\vspace{-15pt}
\caption{Implementation details of CaMDN architecture. (a) NEC handles NPU-specific requests. (b) CPT translates \textit{vcaddr} into \textit{pcaddr} that indexes the targeted cache line in data arrays.}
\vspace{-15pt}
\label{fig_cache}
\end{figure}

\subsection{NPU-controlled Cache Architecture}\label{sec_hardware}
The NPU-controlled cache architecture of CaMDN supports the instantiation of model-exclusive, NPU-controlled regions inside shared cache. Specifically, (1) shared cache is divided into a general-purpose  subspace and an NPU subspace by way partitioning. (2) An NEC is added in each cache slice to take control of the NPU subspace and provide NPU-controlled cache access semantics. (3) The NPU subspace is then divided into model-exclusive regions by paging, which is implemented by a hardware-based CPT inside each NPU.

\subsubsection{Way-Partitioned NPU Subspace}
CaMDN divides shared cache into a general-purpose subspace and an NPU subspace by way partitioning. This eliminates unexpected contention between CPU applications and NPU tasks. CaMDN implements way partitioning by adding a way mask register in each cache slice to mask off the ways allocated to the NPU subspace. As shown in the bottom-right part of Figure \ref{fig_overall_arch}, ways 0-1 are reserved for CPU applications, and ways 2-7 are assigned to the NPU subspace. Different proportions of partitioning can be adapted for different application scenarios. 

\begin{figure*}[t]
\centering
\includegraphics[scale=0.95]{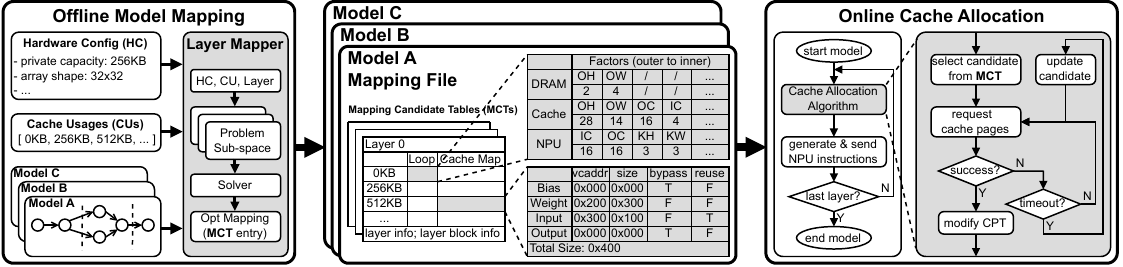}
\vspace{-10pt}
\caption{CaMDN cache scheduling includes offline cache-aware mapping and online cache allocation.}
\vspace{-15pt}
\label{fig_software}
\end{figure*}

\subsubsection{NPU-controlled Access}
To support NPU-controlled cache access, CaMDN adds a lightweight NEC and a dual interface in each cache slice to take control of the NPU subspace, as shown in Figure \ref{fig_cache}(a). The dual interface provides a uniform interface for normal cache requests and NPU-specific requests. NEC handles different types of NPU requests and performs operations including reading/writing cache lines and generating memory requests to memory controllers.

NEC implements NPU-controlled cache access semantics with cache line granularity. Replacing the original hardware-managed cache control, these semantics provide NPU programs with full control over data movement, storage and replacement. In addition to basic semantics which implement memory-cache and cache-NPU data movements, NEC introduces advanced semantics for performance improvement: bypass and multicast. Bypass semantics support bypassing non-reusable data around shared cache, reserving cache space for reusable data to improve utilization. This includes (1) \textbf{bypass-read} a line from memory to NPU; (2) \textbf{bypass-write} a line from NPU to memory. Multicast semantics help reduce bandwidth pressure on both memory and NoC by combining identical requests from a group of NPUs that are allocated to the same model. This includes (3) \textbf{multicast-read} a line from cache to a group of NPUs; (4) \textbf{multicast-bypass-read} a line from memory to a group of NPUs.

\subsubsection{Virtual Cache Addressing}
To further make the NPU subspace divisible and addressable for co-located models, CaMDN utilizes hardware-based paging to create isolated model-exclusive regions and provide an independent virtual address space for every model. As shown in the bottom-right part of Figure \ref{fig_overall_arch}, the NPU subspace is divided into pages of identical size and assigned to different models. A hardware-based CPT is installed in each NPU to translate virtual cache address (\textit{vcaddr}) into physical cache address (\textit{pcaddr}). Considering that NPUs usually operate on data chunks of at least dozens of KB, we set the page size to 32KB for a 16MB shared cache. Thus, a hardware-based CPT has at most 512 entries, each of which needs at most 3 bytes to store a physical cache page number (\textit{pcpn}) and a valid bit, resulting in a total 1.5KB SRAM overhead, which is negligible as demonstrated in Section \ref{eval_overhead}.  

Figure \ref{fig_cache}(b) illustrates the address translation and cache line indexing. The virtual cache page number (\textit{vcpn}) of a \textit{vcaddr} is used to index the CPT for \textit{pcpn}, which is then used to compose \textit{pcaddr}. \textit{pcaddr} is divided into four bit-fields: byte offset, slice index, set index and way index, from lower to higher, which uniquely determine the targeted cache line. In this indexing manner, consecutive data lines are distributed among all slices for higher cache bandwidth utilization.

\subsection{Cache-aware Mapping}\label{sec_mapping}
The cache-aware mapping method of CaMDN provides adaptability to varying available cache capacity by generating multiple mapping candidates that target different cache usage levels. As shown in the left part of Figure \ref{fig_software}, it takes a hardware configuration, a list of cache usage limitations, and model structures as inputs. For each layer, the layer mapper generates a mapping candidate within each usage limitation. All candidates of the layer are written into a mapping candidate table (MCT) and all MCTs of the model are stored in a model mapping file as the output of the mapping phase.

\subsubsection{Heuristic-solver-hybrid Layer Mapper}
To reduce the additional time cost of generating multiple mapping candidates, CaMDN introduces a heuristic-solver-hybrid layer mapper that combines heuristic rules and an available optimization solver to speed up the mapping process while maintaining high performance of results. Specifically, it first shrinks the problem space according to a set of heuristic rules. These rules improve the utilization of cache line, NPU-private storage and compute resource, and reduce the choices of loop permutation\cite{dmazerunner}. Second, it constructs a set of disjoint problem subspaces, each of which is an integer programming problem that takes minimal DRAM access as the optimization objective. Third, these problems are solved by the solver, and the result with the minimal DRAM access is selected as the mapping candidate for the layer within a certain cache usage limitation.

\subsubsection{Layer-block Mapping for Data Reuse}
In addition to layer-wise mapping (LWM) mentioned above, CaMDN also uses layer-block mapping (LBM) to store intermediate data between layers fully in cache and allocate zero memory space to these data. LBM significantly reduces memory access during multi-tenant DNN execution. To prevent a model from occupying too much cache space for too long, models are segmented into layer blocks and LBM happens only inside each block. For each layer, an additional LBM candidate is generated alongside multiple LWM candidates. 

\subsubsection{Mapping Candidate Table}
Mapping candidate tables (MCTs) are the major outputs of the offline mapping phase. An MCT records all mapping candidates and detailed information of a layer in a compact format instead of unrolled NPU instructions. In this way, much less storage is needed for multiple mapping candidates. As shown in the middle of Figure \ref{fig_software}, for each candidate, there is a loop table that records loop permutations and factors, and a cache map table that indicates how tensors are mapped in vcaddr space. 

\begin{algorithm}[t]
\caption{Predict near-future shared cache usage and select mapping candidate accordingly for a layer.}
\label{algo_select}

\fontsize{9}{11}\selectfont

\KwIn{$\rm t_{cur}$: current task with its model information; \newline $\rm MCT_{cur}$: MCT for current layer of $\rm t_{cur}$.}
\KwOut{$\rm M_{cur}$: mapping candidate selected for $\rm t_{cur}$; \newline $\rm P_{cur}$: shared cache pages required by $\rm M_{cur}$; \newline $\rm T_{ahead}$: timeout threshold for waiting.}
\KwData{$\rm T_{next}[numTasks]$: profiling-based predicted next reallocation time for each task; \newline $\rm P_{next}[numTasks]$: predicted pages needed at next reallocation for each task; \newline $\rm P_{alloc}[numTasks]$: allocated pages for each task. }

\tcp{Predict future available pages.}
\SetKwFunction{PAP}{\textbf{Func} predAvailPages}
\SetKwProg{Func}{}{}{}
\Func{\PAP{$\rm T_{ahead}$, $\rm t_{cur}$}$\rm : P_{ahead}$}{
    $\rm P_{ahead} = idlePages()$\;
    \For {$\rm t_{i} : runningTasks$}{
        \If {$\rm t_{i} \ne t_{cur}$ \rm\textbf{and} $\rm T_{next}[t_i] < T_{ahead}$}{
            $\rm P_{ahead}\ +\!\!= P_{alloc}[t_i] - P_{next}[t_i]$\;
        }
    }
    \Return $\rm P_{ahead}$
}

\tcp{Check if LBM is available.}
\If{$\rm hasEnabledLBM(t_{cur})$} {
    $\rm M_{cur} = MCT_{cur}.LBM$\;
    \Return $\rm M_{cur},\ M_{cur}.P_{need},\ \infty$\;
}\ElseIf{$\rm isHeadLayerOfBlock(t_{cur}.layer_{cur})$}{
    $\rm T_{ahead} = T_{cur} + t_{cur}.layerBlock_{cur}.T_{est}\times0.2$\;
    $\rm P_{ahead} = predAvailPages(T_{ahead}, t_{cur})$\;
    \If{$\rm MCT_{cur}.LBM.P_{need} < P_{ahead}$}{
        $\rm M_{cur} = MCT_{cur}.LBM$\;
        \Return $\rm M_{cur},\ M_{cur}.P_{need},\ T_{ahead}$\;
    }
}

\tcp{Select a LWM candidate from MCT.}
$\rm T_{ahead} = T_{cur} + t_{cur}.layer_{cur}.T_{est}\times0.2$\;
$\rm P_{ahead} = predAvailPages(T_{ahead}, t_{cur})$\;
$\rm M_{cur} = MCT_{cur}.LWMs[0]$\;
\For{$\rm M_i : MCT_{cur}.LWMs$}{
    \If{$\rm M_{cur}.P_{need} < M_i.P_{need} \le P_{ahead}$}{
        $\rm M_{cur} = M_i$\;
    }
}
\Return $\rm M_{cur},\ M_{cur}.P_{need},\ T_{ahead}$\;
\end{algorithm}

\subsection{Dynamic Cache Allocation Algorithm}
\label{sec_scheduling}

The dynamic cache allocation algorithm of CaMDN adjusts cache usage among co-located DNNs by predicting near-future cache usage and selecting mapping candidates accordingly.

As shown in the right part of Figure \ref{fig_software}, the algorithm is invoked at the beginning of each layer. First, it predicts the cache usage among tasks in the near future, estimates the available capacity for the layer, and selects the mapping candidate that best fits the available capacity by Algorithm \ref{algo_select}. Second, it tries to request the cache pages needed. If the pages become available within a threshold time, the CPTs are modified and the layer is executed with the selected mapping. Otherwise, every time a timeout occurs, it updates the candidate to the one that require fewer pages.

Algorithm \ref{algo_select} describes the detailed process of predicting shared cache usage and selecting a mapping candidate for a layer. The algorithm uses function $\rm predAvailPages$ to predict the available pages $\rm P_{ahead}$ in future time $\rm T_{ahead}$ for task $\rm t_{cur}$ (lines 1-6). The function is based on three global variables, $\rm T_{next}$, $\rm P_{next}$ and $\rm P_{alloc}$, which are updated at the end of each layer. The algorithm takes the current task $\rm t_{cur}$ and the current layer's MCT $\rm MCT_{cur}$ as inputs, and produces a mapping candidate $\rm M_{cur}$, a number of required pages $\rm P_{cur}$, and a timeout threshold $\rm T_{ahead}$ as outputs. Specifically, it first checks whether LBM is already enabled or can be enabled according to predicted near-future cache usage (lines 7-15). If available, LBM is enabled for this layer block. Otherwise, the LWM candidate that best fits the near-future available capacity is selected (lines 16-22). The timeout threshold $\rm T_{ahead}$ is calculated twice for both steps, using a profiling-based estimation of execution latency $\rm t_{cur}.layer_{cur}.T_{est}$.

\begin{table}
\setlength{\tabcolsep}{3pt}
\centering
\caption{Benchmark Models for Multi-tenant Execution}
\vspace{-5pt}
\begin{tabular}{l|l|c|c|r} 
\hline
\textbf{Domain}                     & \multicolumn{1}{c|}{\textbf{Model}}   & \textbf{Abbr.}    & \textbf{Type}  & \textbf{QoS(ms)} \\ 
\hline\hline
\multirow{4}{*}{Computer Vision}    & ResNet50\cite{resnet}                 & RS.               & Conv           & 6.7 \\ 
\cline{2-5}
                                    & MobileNet-v2\cite{mobilenetv2}        & MB.               & DwConv         & 2.8 \\ 
\cline{2-5}
                                    & EfficientNet-b0\cite{efficientnet}    & EF.               & DwConv         & 2.8 \\ 
\cline{2-5}
                                    & ViT-base-16\cite{vit}                 & VT.               & Trans          & 40.0 \\ 
\hline
\multirow{2}{*}{\makecell[l]{Natural Language \\ Processing}} & BERT-base\cite{bert} & BE.      & Trans          & 40.0 \\ 
\cline{2-5}
                                    & GNMT\cite{gnmt}                       & GN.               & LSTM           & 6.7 \\ 
\hline
Audio Processing                    & Wav2Vec2-base\cite{wav2vec2}          & WV.               & Trans          & 16.7\\ 
\hline
Point Cloud                         & PointPillars\cite{pointpillars}       & PP.               & Conv           & 100.0 \\
\hline
\end{tabular}
\label{tab_benchmark}
\end{table}

\begin{table}
\centering
\caption{Configuration of the NPU-Integrated SoC}
\vspace{-5pt}
\begin{tabular}{l|l|r} 
\hline
\multicolumn{2}{c|}{\textbf{Parameter}}               & \multicolumn{1}{c}{\textbf{Value}}  \\ 
\hline\hline
\multirow{3}{*}{NPU}          & PE Array (per core)   & 32x32                               \\ 
\cline{2-3}
                              & Scratchpad (per core) & 256KB                               \\ 
\cline{2-3}
                              & Cores                 & 16                                  \\ 
\hline
\multirow{3}{*}{Shared Cache} & Total Capacity        & 16MB                                \\ 
\cline{2-3}
                              & NPU Ways / Total Ways & 12 / 16                             \\ 
\cline{2-3}
                              & Slices                & 8                                   \\ 
\hline
\multirow{2}{*}{DRAM}         & Total Bandwidth       & 102.4GB/s                           \\ 
\cline{2-3}
                              & Channels              & 4                                   \\
\hline
\multicolumn{2}{l|}{Frequency}                        & 1GHz                                \\
\hline
\end{tabular}
\vspace{-15pt}
\label{tab_hardware}
\end{table}

\section{Evaluation}

\subsection{Experimental Setup}\label{ref_expr_setup}

\subsubsection{Implementation} The architecture of CaMDN is implemented based on Gemmini\cite{gemmini} in Verilog HDL, and synthesized by Synopsys Design Compiler in a 45nm process technology. OpenRAM\cite{openram} is used to generate SRAM macros. For performance evaluation, an in-door cycle-accurate simulator built on DRAMsim3\cite{dramsim3} is used, which is capable to simulate multi-tenant DNNs running on multiple NPUs with a sliced shared cache. 

\subsubsection{Benchmark} Table \ref{tab_benchmark} shows the eight models chosen to compose the multi-tenant benchmark that feature various domains (CV, NLP, audio processing and point cloud) and different model types (Convolution, Depth-wise Convolution, Transformer and LSTM).

\subsubsection{Baseline} MoCA\cite{moca} and AuRORA\cite{aurora} are chosen as baseline: MoCA dynamically schedules memory bandwidth; AuRORA dynamically schedules NPUs and memory bandwidth. For fairness, MoCA and AuRORA are implemented with the same hardware configuration as CaMDN. Two versions of CaMDN are set for comparison: CaMDN(HW-only) equally allocates cache capacity among NPUs without dynamic cache scheduling; CaMDN(Full) refers to the full system of the architecture-scheduling co-design.

\subsubsection{Experiments} To fully understand the performance improvement, we evaluate CaMDN from three aspects: speedup, scaling and quality-of-service(QoS).

\textbf{Speedup} experiment is conducted with configuration in Table \ref{tab_hardware}. we randomly dispatch each model task to one NPU as soon as it finishes its current task. All NPUs are kept busy to maximize the shared cache contention. 

\textbf{Scaling} experiment is similar to speedup experiment, but with different cache sizes from 4MB to 64MB, and different numbers of co-located DNNs from 1 to 16.

\textbf{QoS} experiment is conducted with configuration in Table \ref{tab_hardware}. CaMDN is integrated with the same memory bandwidth and NPU allocation algorithms as AuRORA. We set tighter latency targets than prior work, as shown in Table \ref{tab_benchmark}. Following \cite{aurora}, we setup three different QoS levels: QoS-H, QoS-M and QoS-L, respectively referring to 0.8$\times$, 1.0$\times$ and 1.2$\times$ latency targets. The following metrics are used in comparison: Service Level Agreement (SLA) satisfaction rate indicates the percentage of tasks that meet their deadline; System Throughput (STP) indicates the overall throughput; Fairness indicates the equality degree of the progresses of co-running tasks. Detailed definitions are described in \cite{aurora}.

\begin{figure}[t]
\centering
\includegraphics[scale=0.49]{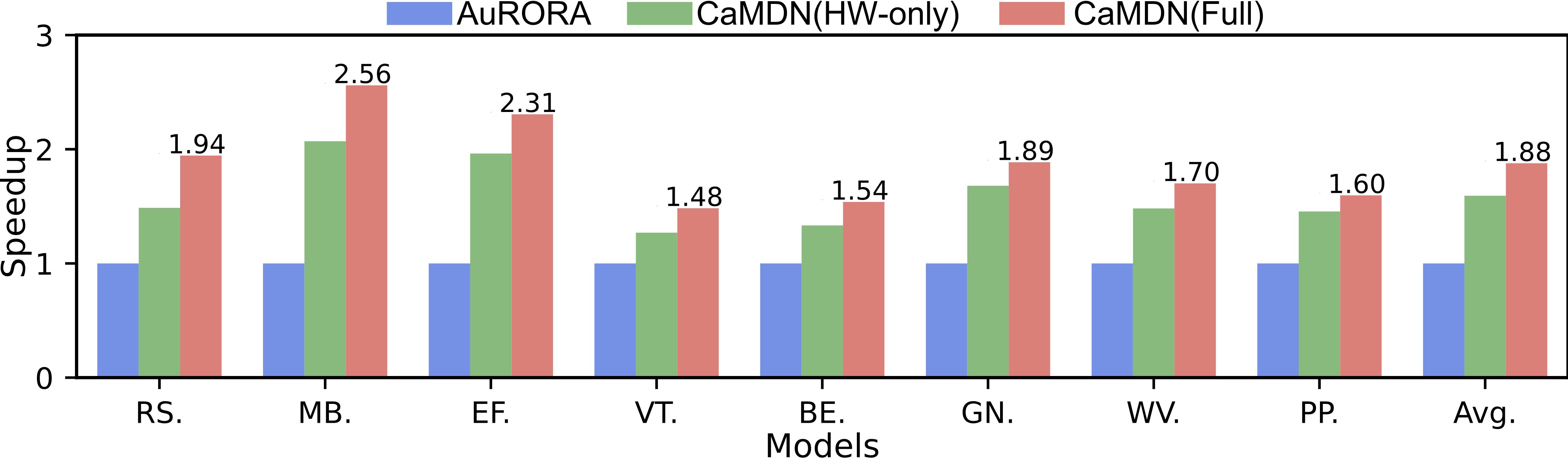}
\vspace{-5pt}
\caption{The model-wise speedup achieved by CaMDN.}
\vspace{-10pt}
\label{fig_eval_speedup}
\end{figure}

\begin{figure}[!t]
\centering
\includegraphics[scale=0.49]{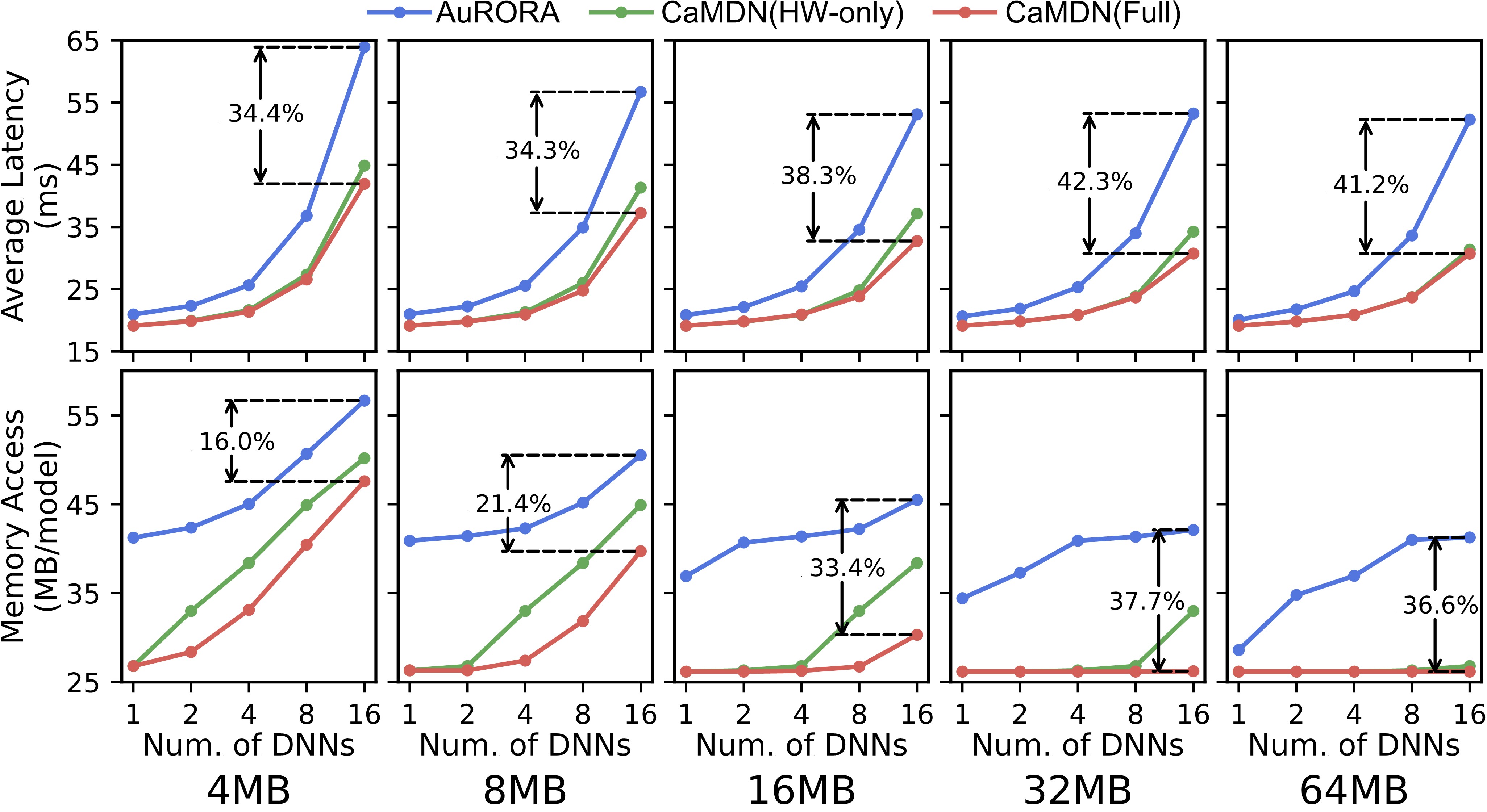}
\vspace{-5pt}
\caption{The average latency and memory access with different scales. The numbers indicate the percentage reduction achieved by CaMDN(Full) compared to the baseline.}
\vspace{-20pt}
\label{fig_eval_scaling}
\end{figure}

\vspace{-3pt}
\subsection{Results Analysis}
\vspace{-3pt}

\subsubsection{Speedup}
Figure \ref{fig_eval_speedup} shows the model-wise speedup of CaMDN. Since the methods in MoCA and AuRORA are essentially for improving QoS rather than speedup, and they show similar results in this experiment, we choose AuRORA as a representative. CaMDN(Full) achieves an averagely 1.88$\times$ speedup, demonstrating that enhancing cache efficiency significantly improves the model speed. CaMDN(Full) surpasses CaMDN(HW-only) by an averagely 1.18$\times$ speedup, proving the effectiveness of dynamic cache allocation. CaMDN reaches higher speedup on MobileNet-v2 and EfficientNet-b0 because these models possess larger proportions of intermediate data and CaMDN successfully reduces the memory access by storing intermediate data entirely in cache.  

\subsubsection{Scaling}
Figure \ref{fig_eval_scaling} shows the average model latency and memory access with different system scales. Generally, CaMDN(Full) achieves 34.3\% to 42.3\% latency reductions and 16.0\% to 37.7\% memory access reductions. Since multi-tenant DNNs are memory-bound, the reduction in memory access contributes a significant portion of performance improvement. As the number of co-located DNNs increases, CaMDN significantly alleviates the performance degradation. As the cache becomes larger, CaMDN generally shows larger enhancement. 

\subsubsection{QoS}
\begin{figure}[t]
\centering~
\includegraphics[scale=0.49]{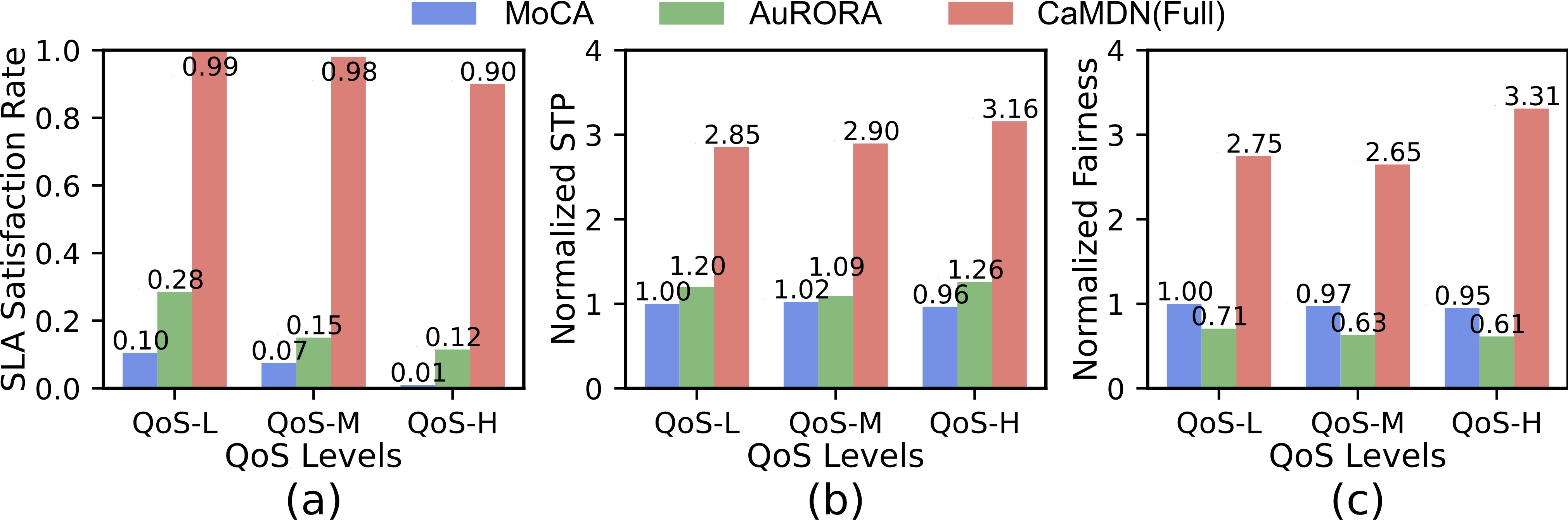}
\vspace{-5pt}
\caption{The QoS improvement achieved by CaMDN.}
\vspace{-5pt}
\label{fig_eval_qos}
\end{figure}

As shown in Figure \ref{fig_eval_qos}, CaMDN achieves averagely 5.9$\times$, 2.5$\times$ and 3.0$\times$ improvements on SLA satisfaction rate, STP and fairness respectively. This demonstrates that CaMDN can meet more rigorous QoS requirements by improving cache efficiency to reduce the model latency. Notably, CaMDN achieves even higher STP and fairness in QoS-H. The reason is that bandwidth and NPUs are more frequently allocated to slow tasks in QoS-H, while in QoS-L and QoS-M, the improvement of CaMDN largely reduces the need for bandwidth and NPU allocation. Additionally, AuRORA shows lower fairness than MoCA in our experiment, because our QoS targets are much tighter and AuRORA manages to reach higher SLA satisfaction rate at the cost of worsening fairness. Nonetheless, CaMDN can achieve higher performance without sacrificing fairness.

\subsubsection{Area Overhead}\label{eval_overhead}
Table \ref{tab_overhead} shows the negligible area overhead of an NPU and a cache slice with the configuration in Table \ref{tab_hardware}. A CPT contributes 0.9\% of the total NPU area. An NEC contributes 0.3\% of the total cache slice area.

\begin{table}[t]
\setlength{\tabcolsep}{3pt}
\centering
\caption{Area Breakdown of the Architecture of CaMDN}
\vspace{-5pt}
\begin{tabular}{l|c|cll|c|c} 
\cline{1-3}\cline{5-7}
\textbf{Component}  & \textbf{Area($\mu m^2$)}  & \textbf{(\%)} & & \textbf{Component}  & \textbf{Area($\mu m^2$)}  & \textbf{(\%)}      \\ 
\hhline{===~===}
NPU                 & 7905k                     & 100.0         & & Cache Slice         & 24676k                    & 100.0                  \\
~ Scatchpad         & 6302k                     & 79.7          & & ~ Data Array        & 21878k                    & 88.7                   \\
~ PE Array          & 1302k                     & 16.5          & & ~ Tag Array         & 2398k                     & 9.7                    \\
~ \textbf{CPT}      & \textbf{73k}              & \textbf{0.9}  & & ~ \textbf{NEC}      & \textbf{66k}              & \textbf{0.3}  \\
~ others            & 228k                      & 2.9           & & ~ others            & 334k                      & 1.3                    \\
\cline{1-3}\cline{5-7}
\end{tabular}
\vspace{-10pt}
\label{tab_overhead}
\end{table}

\section{Conclusion}

We proposed CaMDN, an architecture-scheduling co-design to enhance cache efficiency for multi-tenant DNNs on integrated NPUs. Compared to prior works that focus on memory bandwidth and/or compute resources, CaMDN is dedicated to solving the inefficiency issue of cache shared by multiple NPUs. CaMDN combines a lightweight architecture, a cache-aware mapping method and a dynamic cache allocation algorithm. CaMDN reduces the memory access by 33.4\% and achieves a model speedup of up to 2.56$\times$ (1.88$\times$ on average). For future work, we believe more efficient scheduling methods could be proposed that take both multi-tenant DNNs and general-purposed programs into consideration.

\clearpage
\bibliographystyle{IEEEtran}
\bibliography{ref}

\end{document}